\documentstyle[prd,preprint,tighten,aps,amssymb,newlfont,epsfig]{revtex}
\begin{document}
\draft
\preprint{
\begin{tabular}{r}
   UWThPh-1999-24
\\ DFTT 22/99
\\ hep-ph/9904316
\end{tabular}
}
\title{Four-neutrino spectrum from oscillation data}
\author{S.M. Bilenky}
\address{Joint Institute for Nuclear Research, Dubna, Russia, and\\
Institute for Theoretical Physics, University of Vienna,\\
Boltzmanngasse 5, A--1090 Vienna, Austria}
\author{C. Giunti}
\address{INFN, Sez. di Torino, and Dip. di Fisica Teorica,
Univ. di Torino,\\
Via P. Giuria 1, I--10125 Torino, Italy}
\author{W. Grimus and T. Schwetz}
\address{Institute for Theoretical Physics, University of Vienna,\\
Boltzmanngasse 5, A--1090 Vienna, Austria}
\maketitle
\begin{abstract}
It is shown that the Super-Kamiokande atmospheric up--down asymmetry,
together with the results of all other neutrino oscillation experiments,
allows to constraint the possible spectra of four massive neutrinos.
The two schemes with two pairs of neutrinos with close masses 
separated by a gap of about 1 eV are favored by the data.
\end{abstract}

\pacs{Talk presented by C. Giunti at the
17$^{\mathrm{th}}$ \textit{International Workshop on Weak Interactions and Neutrinos}
(WIN99),
Cape Town, South Africa, 24--30 January 1999.}

The recent observation of an up--down asymmetry of high-energy
$\mu$-like events generated by atmospheric neutrinos in the
Super-Kamiokande experiment\cite{SK-atm-98}
represents a convincing
model-independent evidence in favor of neutrino oscillations.
Indications in favor of disappearance of atmospheric $\nu_\mu$'s
have been obtained also in
the Kamiokande and IMB experiments
and in the recent
Soudan 2 and MACRO experiments.\cite{BGG98-review}
Other indications in favor of neutrino oscillations have been obtained in
solar neutrino experiments
(Homestake,
Kamiokande,
GALLEX,
SAGE,
Super-Kamiokande)
and in the LSND experiment.\cite{BGG98-review}
The flux of electron neutrinos measured in
all five solar neutrino experiments
is substantially smaller than the one predicted
by the Standard Solar Model
and a comparison of the data of different experiments
indicate an energy dependence of the solar $\nu_e$ suppression,
which represents a rather convincing evidence
in favor of neutrino oscillations.\cite{BGG98-review}
The accelerator LSND experiment
is the only one that claims the observation
of neutrino oscillations in appearance channels,
specifically $\bar\nu_\mu\to\bar\nu_e$ and $\nu_\mu\to\nu_e$.

The probabilities of neutrino oscillations
depend on the elements of the neutrino mixing matrix
$U$,
that connects the flavor neutrino fields $\nu_{\alpha L}$
to the massive neutrino fields $\nu_{kL}$
through the relation
$ \nu_{\alpha L} = \sum_k U_{\alpha k} \nu_{kL} $,
and on the phases
$ \Delta{m}^2_{kj} L / E $,
where
$\Delta{m}^2_{kj} \equiv m_k^2 - m_j^2$
($m_k$ is the mass of the neutrino field $\nu_k$),
$L$ is the source-detector distance
and $E$ is the neutrino energy.
If
$ \Delta{m}^2_{kj} L / E \ll 1 $
neutrino flavor transitions cannot be observed
and
if
$ \Delta{m}^2_{kj} L / E \gg 1 $
only the averaged transition probability
can be measured.
Since
a variation of the transition probability as a function of neutrino energy
has been observed in all the experiments mentioned above
and the range of $L/E$ probed by each type of experiment is different
($ L / E \gtrsim 10^{10} \, \mathrm{eV}^{-2} $
for solar neutrino experiments,
$ L / E \sim 10^{2} - 10^{3} \, \mathrm{eV}^{-2} $
for atmospheric neutrino experiments
and
$ L / E \sim 1 \, \mathrm{eV}^{-2} $
for the LSND experiment),
it is clear that in order to explain all the observations with neutrino oscillations
at least three $\Delta{m}^2$'s with different scales are needed:
$ \Delta{m}^2_{\mathrm{sun}} \lesssim 10^{-10} \, \mathrm{eV}^2 $,
$ \Delta{m}^2_{\mathrm{atm}} \sim 10^{-3} - 10^{-2} \, \mathrm{eV}^2 $,
$ \Delta{m}^2_{\mathrm{LSND}} \sim 1 \, \mathrm{eV}^2 $
(if the MSW effect\cite{BGG98-review}
is responsible of solar neutrino transitions,
$ \Delta{m}^2_{\mathrm{sun}} $
must be smaller than about
$ 10^{-4} \, \mathrm{eV}^2 $
in order to have a resonance in the interior of the sun
and is still at least one order of magnitude smaller than
$\Delta{m}^2_{\mathrm{atm}}$).
This means that at least four light massive neutrinos must exist in nature.
Here we consider the minimal possibility of four neutrinos,
which implies that the three flavor neutrinos $\nu_e$, $\nu_\mu$, $\nu_\tau$
are accompanied by a sterile neutrino $\nu_s$
that does not take part in
standard weak interactions.

\begin{figure}[h]
\begin{center}
\epsfig{file=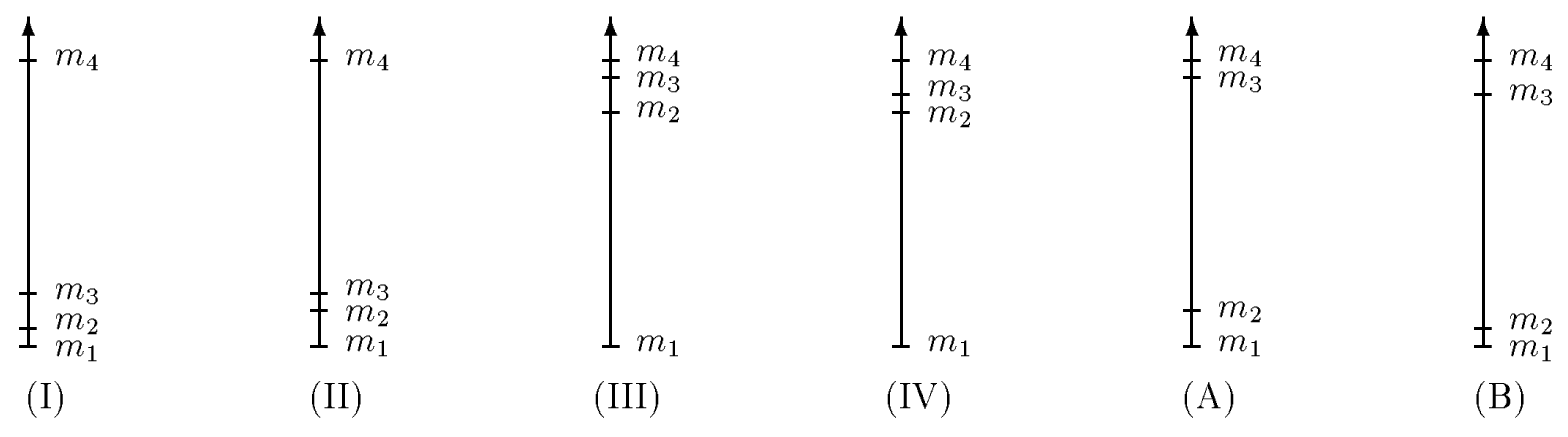,width=0.99\linewidth}
\caption{ \label{4spectra} }
\end{center}
\end{figure}

The six types of four-neutrino mass spectra that can accommodate
the hierarchy
$
\Delta{m}^2_{\mathrm{sun}}
\ll
\Delta{m}^2_{\mathrm{atm}}
\ll
\Delta{m}^2_{\mathrm{LSND}}
$
are shown in Fig.~\ref{4spectra}.
In all these mass spectra there are two groups
of close masses separated by the ``LSND gap'' of the order of 1 eV.
In each scheme the smallest mass-squared
difference corresponds to
$\Delta{m}^2_{\mathrm{sun}}$
($\Delta{m}^2_{21}$ in schemes I and B,
$\Delta{m}^2_{32}$ in schemes II and IV,
$\Delta{m}^2_{43}$ in schemes III and A),
the intermediate one to
$\Delta{m}^2_{\mathrm{atm}}$
($\Delta{m}^2_{31}$ in schemes I and II,
$\Delta{m}^2_{42}$ in schemes III and IV,
$\Delta{m}^2_{21}$ in scheme A,
$\Delta{m}^2_{43}$ in scheme B)
and the largest mass squared difference
$ \Delta{m}^2_{41} = \Delta{m}^2_{\mathrm{LSND}} $
is relevant for the oscillations observed in the LSND experiment.

It has been shown\cite{BGG96} that
the schemes I--IV are disfavored by the
results of short-baseline accelerator and reactor disappearance
neutrino oscillation experiments for all values of $\Delta{m}^2_{41}$
in the LSND-allowed range
0.2 -- 2 eV$^2$,
with the possible exception of the small interval from
0.2 to 0.3 eV$^2$
where there are no data from $\nu_\mu$
short-baseline disappearance experiments.
Here we will show\cite{BGGS99} that this gap is closed
by the inclusion in the analysis of
the asymmetry of $\mu$-like high-energy events
\begin{equation}
\mathcal{A} = (U-D)/(U+D) = - 0.311 \pm 0.043 \pm 0.01
\label{A}
\end{equation}
measured in the Super-Kamiokande experiment.\cite{SK-atm-98}
Here $U$ and $D$ are the number of events in the zenith angle
intervals $-1 < \cos \theta < -0.2$ and $0.2 < \cos \theta < 1$,
respectively.

In the following we will consider only the scheme I with a mass hierarchy,
but the results
apply also to the schemes II, III and IV.
Let us remark that in principle one could check which scheme is allowed
by doing a combined fit of all data.
However,
at the moment it is not possible to perform such a fit
because of the enormous complications due
to the presence of many parameters (six mixing angles, etc.)
and to the difficulties involved in a combined fit of
the data of different experiments,
which are usually analyzed by the experimental collaborations
using different methods.
Hence,
we think that it is quite remarkable
that one can exclude the schemes I--IV
with the following relatively simple procedure.

The exclusion plots obtained in short-baseline
$\bar\nu_e$ and $\nu_\mu$
disappearance experiments
imply that\cite{BGG96}
$
|U_{\alpha4}|^2 \leq a^0_\alpha
\quad \mbox{or} \quad
|U_{\alpha4}|^2 \geq 1-a^0_\alpha
$
for $\alpha=e,\mu$,
with\cite{BGG98-review}
$ a_e^0 \lesssim 4 \times 10^{-2} $
for
$\Delta m^2_{41} \gtrsim 0.1 \, \mathrm{eV}^2$
and
$ a_\mu^0 \lesssim 0.2 $
for
$\Delta m^2_{41} \gtrsim 0.4 \, \mathrm{eV}^2$.
However,
since the survival probability of solar $\nu_e$'s is bounded by\cite{BGG96} 
$
P^{\mathrm{sun}}_{\nu_e\to\nu_e} \geq |U_{e4}|^4
$,
only the range
\begin{equation}
|U_{e4}|^2 \leq a^0_e
\label{ue4}
\end{equation}
is acceptable.
In a similar way,
since the survival probability of atmospheric $\nu_\mu$'s and $\bar\nu_\mu$'s
(all the following inequalities are valid
both for neutrinos and antineutrinos)
is bounded by\cite{BGG96} 
$
P^{\mathrm{atm}}_{\nu_\mu\to\nu_\mu} \geq |U_{\mu4}|^4
$,
it is clear that large values of
$|U_{\mu4}|^2$
are incompatible with the observed asymmetry (\ref{A}).

\begin{figure}[h]
\begin{center}
\epsfig{file=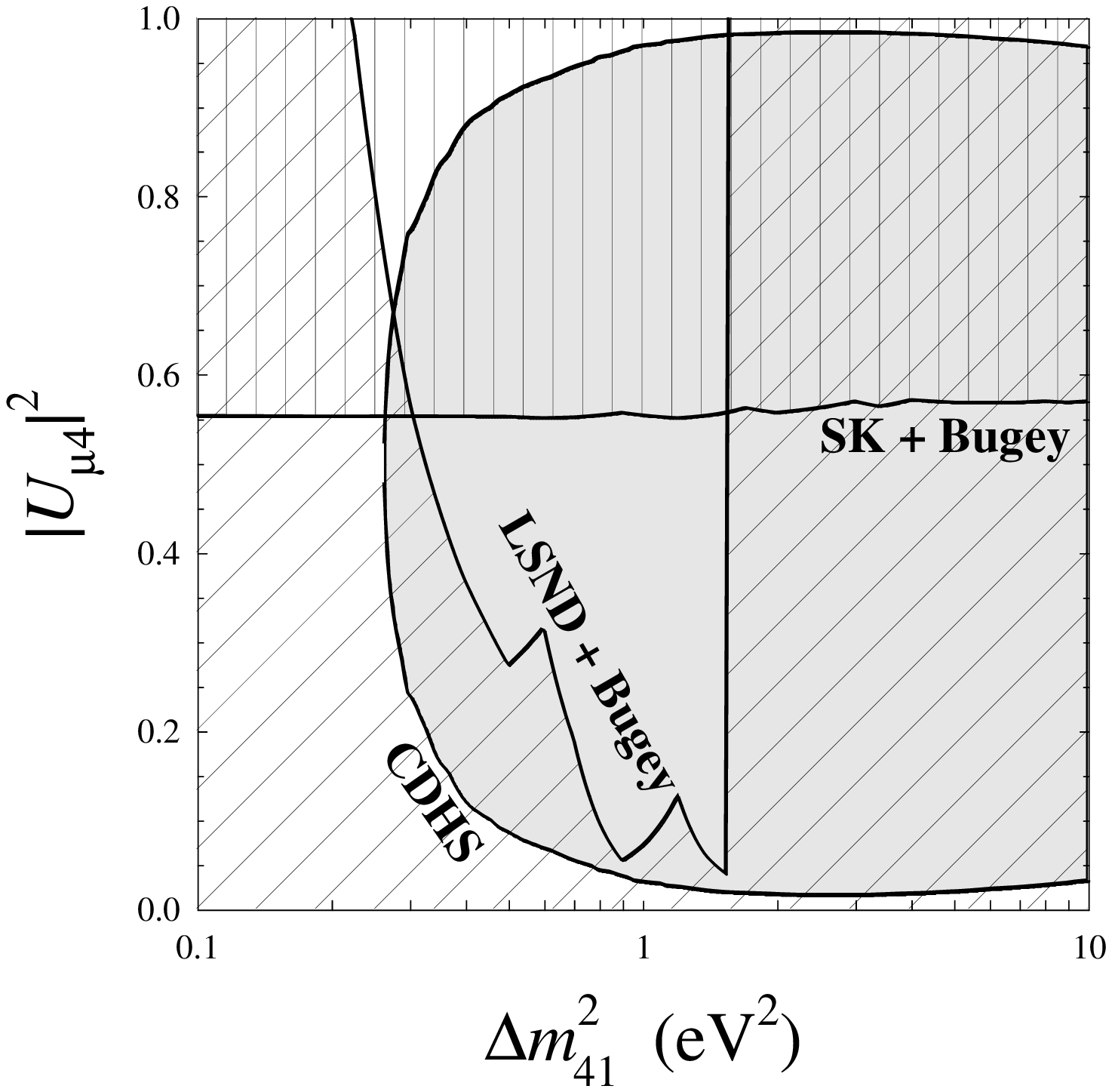,width=0.50\linewidth}
\caption{ \label{umu4} }
\end{center}
\end{figure}

Let us derive the upper bound for
$|U_{\mu4}|^2$
that follows from the asymmetry (\ref{A}).
Because of the small value of 
$\Delta m^2_{\mathrm{atm}} = \Delta m^2_{31}$,
downward-going neutrinos
do not oscillate with the atmospheric
mass-squared difference and the
survival probability of downward-going neutrinos given by
\begin{equation}
P^D_{\nu_\alpha\to\nu_\alpha}
=
|U_{\alpha4}|^4 + \left( 1 - |U_{\alpha4}|^2 \right)^2
\,.
\label{PD}
\end{equation}
The conservation of probability and Eq.(\ref{ue4})
allow to deduce the upper bound
\begin{equation}
P^D_{\nu_e\to\nu_\mu} \leq 1 - P^D_{\nu_e\to\nu_e} =
2 \, |U_{e4}|^2 \left( 1 - |U_{e4}|^2 \right)
\leq
2\, a^0_e (1-a^0_e)
\,.
\label{bemu}
\end{equation}
From Eqs.(\ref{PD}) and (\ref{bemu}) we obtain the upper bound
\begin{equation}
D
\leq
N_\mu \left[ |U_{\mu4}|^4 + \left( 1 - |U_{\mu4}|^2 \right)^2 \right]
+ 2 \, N_e a^0_e (1-a^0_e)
\,,
\label{Du}
\end{equation}
where
$N_\mu$ and $N_e$ are
number of muon (electron) neutrinos and antineutrinos 
produced in the atmosphere.
On the other hand,
taking into account only the part of $D$ which is determined by the
survival probability of $\nu_\mu$'s, we obtain the lower bound
\begin{equation}\label{Dl}
D
\geq
N_\mu \left[ |U_{\mu4}|^4 + \left( 1 - |U_{\mu4}|^2 \right)^2 \right]
\,.
\end{equation}
Furthermore,
using the lower bound\cite{BGG96} 
$
P^{\mathrm{atm}}_{\nu_\mu\to\nu_\mu} \geq |U_{\mu4}|^4
$,
for upward-going neutrinos we have
\begin{equation}\label{Ul}
U \geq N_\mu \, |U_{\mu4}|^4
\,.
\end{equation}
With the inequalities (\ref{Du}), (\ref{Dl}) and (\ref{Ul}),
for the asymmetry (\ref{A}) we obtain
\begin{equation}\label{ineq}
- \mathcal{A}
\leq
\frac
{ \left( 1 - |U_{\mu4}|^2 \right)^2 + 2 \, a^0_e (1-a^0_e) / r }
{ \left( 1 - |U_{\mu4}|^2 \right)^2 + 2 \, |U_{\mu4}|^4 }
\,,
\end{equation}
where $r \equiv N_\mu/N_e \simeq 2.8$.
Solving the inequality (\ref{ineq}) for $|U_{\mu4}|^2$,
we finally obtain the upper bound
\begin{equation}
|U_{\mu4}|^2
\leq
\frac
{ 1 + \mathcal{A}
-
\sqrt{ - 2 \left[ \mathcal{A} ( 1 + \mathcal{A} )
+ ( 1 + 3 \mathcal{A} ) a^0_e (1-a^0_e) / r
\right] } }
{ 1 + 3 \mathcal{A} }
\equiv
a^{\mathrm{SK}}_\mu
\,.
\label{bound}
\end{equation}
Since the measured value (\ref{A}) of $\mathcal{A}$ implies that
$- \mathcal{A} \geq 0.254$ at 90\% CL,
from the inequality (\ref{ineq}) we obtain the upper bound
depicted by the horizontal line in Fig.~\ref{umu4}
(the vertically hatched area is excluded).

In Fig.~\ref{umu4} we have also shown the bound
$|U_{\mu4}|^2 \leq a^0_\mu$
or
$|U_{\mu4}|^2 \geq 1-a^0_\mu$
obtained from the exclusion plot of the short-baseline
CDHS
$\nu_\mu$ disappearance experiment,
which exclude the shadowed region.

The results of the LSND experiment
imply a lower bound
$A^\mathrm{min}_{\mu;e}$
for the amplitude
$A_{\mu;e} = 4 |U_{e4}|^2 |U_{\mu4}|^2$
of $\nu_\mu\to\nu_e$ oscillations,
from which we obtain the constraint
\begin{equation}\label{LSND}
|U_{\mu4}|^2 \geq A^\mathrm{min}_{\mu;e}/4a^0_e
\,.
\end{equation}
This bound is represented by the curve in Fig.~\ref{umu4} labelled
LSND + Bugey
(the diagonally hatched area is excluded). 

From Fig.~\ref{umu4}
one can see that,
in the framework of scheme I,
there is no range of $|U_{\mu4}|^2$
that is compatible with all the experimental data.
Hence,
the scheme with four neutrinos and a mass hierarchy is
strongly disfavored.

\begin{figure}[h]
\begin{center}
\epsfig{file=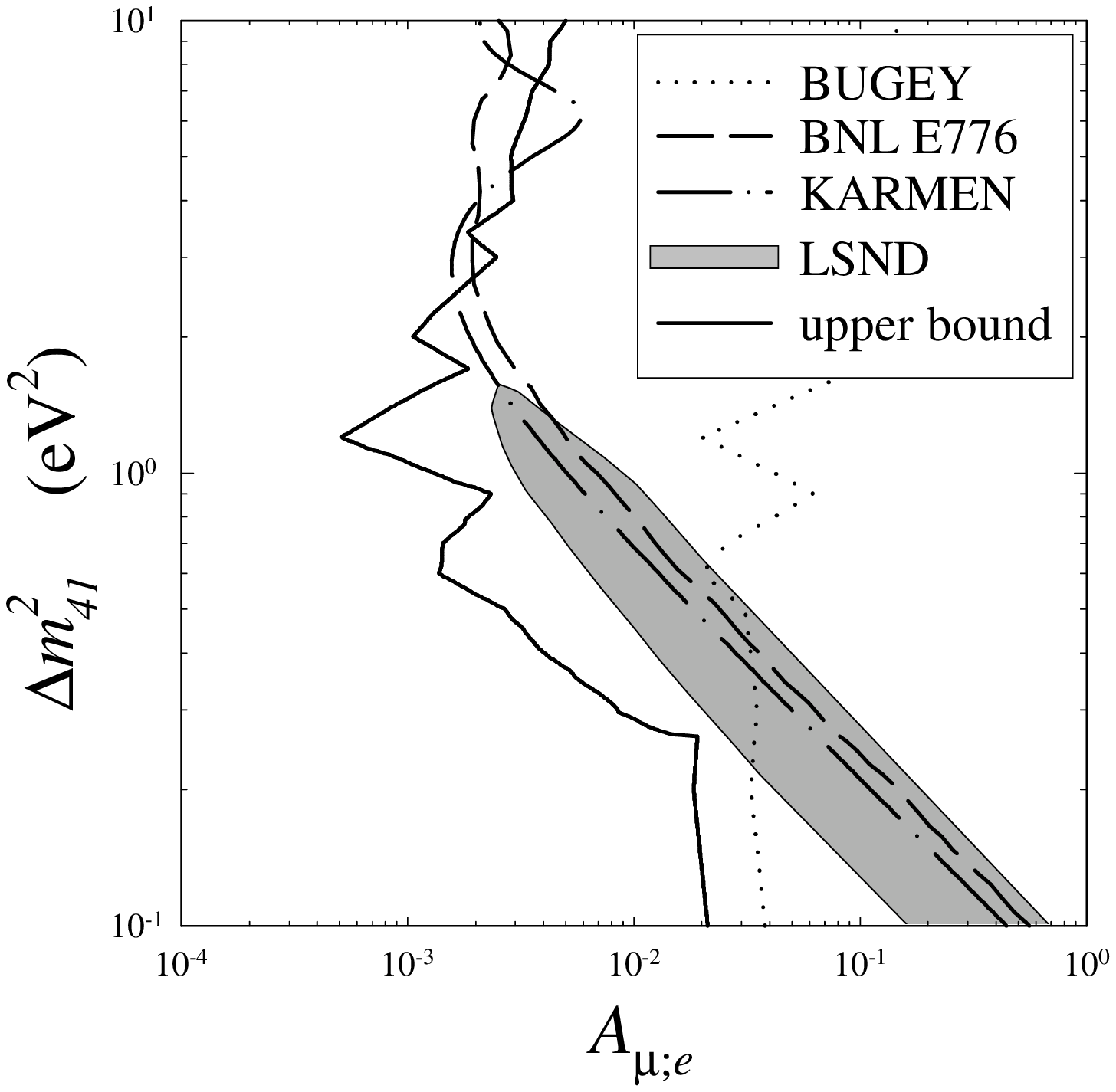,width=0.50\linewidth}
\caption{ \label{amuel} }
\end{center}
\end{figure}

The incompatibility of the experimental results with the mass spectrum I
is shown also in Fig.~\ref{amuel},
where we have plotted in the $A_{\mu;e}$--$\Delta{m}^2_{41}$ plane
the upper bound
$ A_{\mu;e} \leq 4 \, a^0_e \, a^0_\mu $
for
$ \Delta{m}^2_{41} > 0.26 \, \mathrm{eV}^2 $
and
$ A_{\mu;e} \leq 4 \, a^0_e \, a^{\mathrm{SK}}_\mu $
for
$ \Delta{m}^2_{41} < 0.26 \, \mathrm{eV}^2 $
(solid line, the region on the right is excluded).
One can see that this constraint is incompatible with
the LSND-allowed region
(shadowed area).

The procedure presented above for the scheme I applies
also to the schemes II, III and IV,
in which there is a group of three close neutrino
masses separated from the fourth mass
by the LSND gap.
Hence,
we conclude that these schemes are disfavored.
Only the four-neutrino schemes A and B in Fig.~\ref{4spectra}
are compatible with the results of all neutrino oscillation experiments.

\newpage

\end{document}